\documentclass[twocolumn,showpacs,preprintnumbers,showkeys,superscriptaddress,prc]{revtex4}
\usepackage{CJK}
\usepackage{mathrsfs}
\usepackage{amssymb}
\usepackage{amsmath}
\usepackage{graphicx}
\usepackage{dcolumn}
\usepackage{bm}
\usepackage{graphicx}
\usepackage{float}
\usepackage[normalem]{ulem}
\usepackage{color}
\usepackage{multirow}

\newcolumntype{d}[1]{Dc{.}{.}{#1}}
\begin{document}
\begin{CJK*}{UTF8}{}

\title{From local correlations to regional systematics}
\author{Z. Z. Qin ({\CJKfamily{gbsn}覃珍珍})}
\affiliation{School of Science, Southwest University of Science and Technology, Mianyang 621010, China}
\author{M. Bao ({\CJKfamily{gbsn}鲍曼})}
\affiliation{School of Physics and Astronomy, Shanghai Jiao Tong University, Shanghai 200240, China}
\author{Y. Lei ({\CJKfamily{gbsn}雷杨})\footnote{corresponding author: leiyang19850228@gmail.com}}
\affiliation{Key Laboratory of Neutron Physics, Institute of Nuclear Physics and Chemistry, China Academy of Engineering Physics, Mianyang 621900, China}
\date{\today}

\begin{abstract}
Local correlations of $2^+_1$ excitation energies and B(E2, $2^+_1\rightarrow \text{ g.s.}$) values require linear $N_pN_n$ systematics in a logarithmic scale, as confirmed by an experiment survey. Based on local correlations of $\alpha$-decay energies, neutron separation energies, and proton separation energies, one can decouple them into their proton and neutron contributions separately. These contributions exhibit smooth regional systematics beyond the $N_pN_n$ scheme.
\end{abstract}
\maketitle
\end{CJK*}
\section{Introduction}
It is known that $2^+_1$ excitation energies (denoted by $E2$) and B(E2, $2^+_1\rightarrow \text{ g.s.}$) values [denoted by $B(E2)$] of heavy nuclei can be systematized by their $N_pN_n$ product, where $N_p$ and $N_n$ are the numbers of valence protons and neutrons (or holes), respectively \cite{casten-1,casten-2,casten-3}. Recently, local correlations of $E2$ and $B(E2)$ as
\begin{equation}\label{local}
\begin{aligned}
F(N_p,N_n)&+F(N_p+i,N_n+j)-F(N_p+i,N_n)\\
&-F(N_p,N_n+j)\simeq 0,
\end{aligned}
\end{equation}
was empirically suggested to be a generalization of the $N_pN_n$ scheme \cite{bao-npnn-local}, where $F$ refers to the nuclear observable under investigation; $i$ and $j$ can take the value of 1 or 2. Actually, such $E2$ and $B(E2)$ local correlations were already found in the 70s through schematic Hartree-Fock and collective-model derivations \cite{pps}. Later, they were verified by an experimental survey \cite{beta-npnn}, and successfully applied for data prediction \cite{rb-pps-pre}. We note that $\alpha$-decay energies (denoted by $Q_{\alpha}$), neutron separation energies (denoted by $S_n$) and proton separation energies (denoted by $S_p$) are also regulated by the same local correlations as Eq. (\ref{local}) \cite{bao-mass,chen-mass}. These nuclear-mass-related local correlations were derived from the Garvey-Kelson relations \cite{gk}, the linearity in the evolution of neutron ($S_n$) and proton ($S_p$) separation energies \cite{linear-snsp} and the odd-even cancellation of nuclear binding energies \cite{oe-cancel-1,oe-cancel-2}. Using the AME2012 database \cite{q-mass}, their accuracy was demonstrated \cite{bao-mass,chen-mass}. Thus, we believe that Eq. (\ref{local}) is theoretically and empirically reliable for $E2$, $B(E2)$, $Q_{\alpha}$, $S_n$ and $S_p$ values.

In Ref. \cite{bao-npnn-local}, the local correlation behavior of $E2$ and $B(E2)$ values was derived from the functional continuity of the $N_pN_n$ scheme, which corresponds to a first-order approximation of the function $F$ around $(N_p,N_n)$. If higher-order precision is desired, Eq. (\ref{local}) requires additional constraints on the $N_pN_n$ scheme. In contrast, the $Q_{\alpha}$, $S_n$ and $S_p$ data do not yet have $N_pN_n$ systematics. It is still desirable, however,  to find some regularity in regional $Q_{\alpha}$, $S_n$ and $S_p$ evolution in terms of $N_p$ and $N_n$. Thus, in this work we try to clarify the necessary constraint on the $N_pN_n$ scheme from the high-precision requirement of Eq. (\ref{local}), and to develop a new description of nuclear regional systematics from this relation.

\section{formulism}\label{for}
We take the nuclear observable $F$ as a two-dimensional function of $(N_p,N_n)$ in Eq. (\ref{local}), and expand the function $F$ to second order. In this way, Eq. (\ref{local}) is reduced to a second-order partial differential equation about $N_p$ and $N_n$ according to
\begin{equation}\label{partial}
\frac{\partial^2 F}{\partial N_p\partial N_n}\simeq 0.
\end{equation}

If $F$ is also a smooth one-dimensional function of the product $N_pN_n$, then Eq. (\ref{partial}) becomes
\begin{equation}\label{eq-partial}
\frac{\partial^2 F}{\partial N_p\partial N_n}=\frac{{\rm d}F}{{\rm d}(N_pN_n)}+N_pN_n\frac{{\rm d}^2F}{{\rm d}(N_pN_n)^2}\simeq 0.
\end{equation}
The solution of Eq. (\ref{eq-partial}) is
\begin{equation}\label{eq-linear}
F=c_1\ln(N_pN_n)+c_2,
\end{equation}
where $c_1$ and $c_2$ are arbitrary constants. Eq. (\ref{eq-linear}) indicates that when constrained by the high-precision requirement of Eq. (\ref{local}) the $N_pN_n$ plot should exhibit linearity in the logarithmic scale.

On the other hand, the general solution of Eq. (\ref{partial}) is
\begin{equation}\label{formal}
F(N_p,N_n)=f_p(N_p)+f_n(N_n),
\end{equation}
where $f_p$ and $f_n$ are arbitrary functions of $N_p$ and $N_n$, respectively. Eq. (\ref{formal}) indicates that any nuclear observable bound by Eq. (\ref{local}) can be decoupled into separate proton and neutron contributions. If such an observable varies smoothly across the chart of nuclides, its proton and neutron contributions should also exhibit smooth regional systematics as functions of $N_p$ and $N_n$. It is noteworthy that Eq. (\ref{eq-linear}) is just an example of Eq. (\ref{formal}), given that $\ln(N_pN_n)=\ln N_p+\ln N_n$.

\section{logarithmic-$N_pN_n$ linearity of $E2$ and $B(E2)$ values}\label{sys}
$E2$ and $B(E2)$ values are bound both by Eq. (\ref{local}) and by the $N_pN_n$ scheme. Thus, they provide the best platform to illustrate the constraint of Eq. (\ref{local}) on the $N_pN_n$ scheme, {\it i.e.}, the logarithmic linearity defined by Eq. (\ref{eq-linear}). In Figs. \ref{fig-e2} and \ref{fig-be2} we plot the experimental $E2$ and $B(E2)$ values \cite{ensdf,be2-raman} against the logarithmically-scaled $N_pN_n$ product for all six of the major regions of $A>100$ nuclei, partitioned by magic numbers: 28, 50, 82, 126 and 184. Table \ref{tb-region} specifies the range of proton and neutron numbers for each mass region.

\begin{table}
\caption{The range of proton and neutron numbers for the six mass regions under investigation in this work.}\label{tb-region}
\begin{tabular}{ccccccc}
\hline\hline
       &    $Z$ &    $N$       &    $~~~$ &           &    $Z$ &    $N$       \\
\hline
$A\simeq 120$ &    39$\sim$50       &    66$\sim$82       &           &    $A\simeq 170$ &    66$\sim$82       &    82$\sim$104    \\
$A\simeq 130$ &    50$\sim$66       &    66$\sim$82       &           &    $A\simeq 190$ &    66$\sim$82       &    104$\sim$126  \\
$A\simeq 150$ &    50$\sim$66       &    82$\sim$104    &           &    $A\simeq 230$ &    82$\sim$104    &    126$\sim$155  \\
\hline\hline
\end{tabular}
\end{table}

\subsection{$E2$}
\begin{figure}
\includegraphics[angle=0,width=0.48\textwidth]{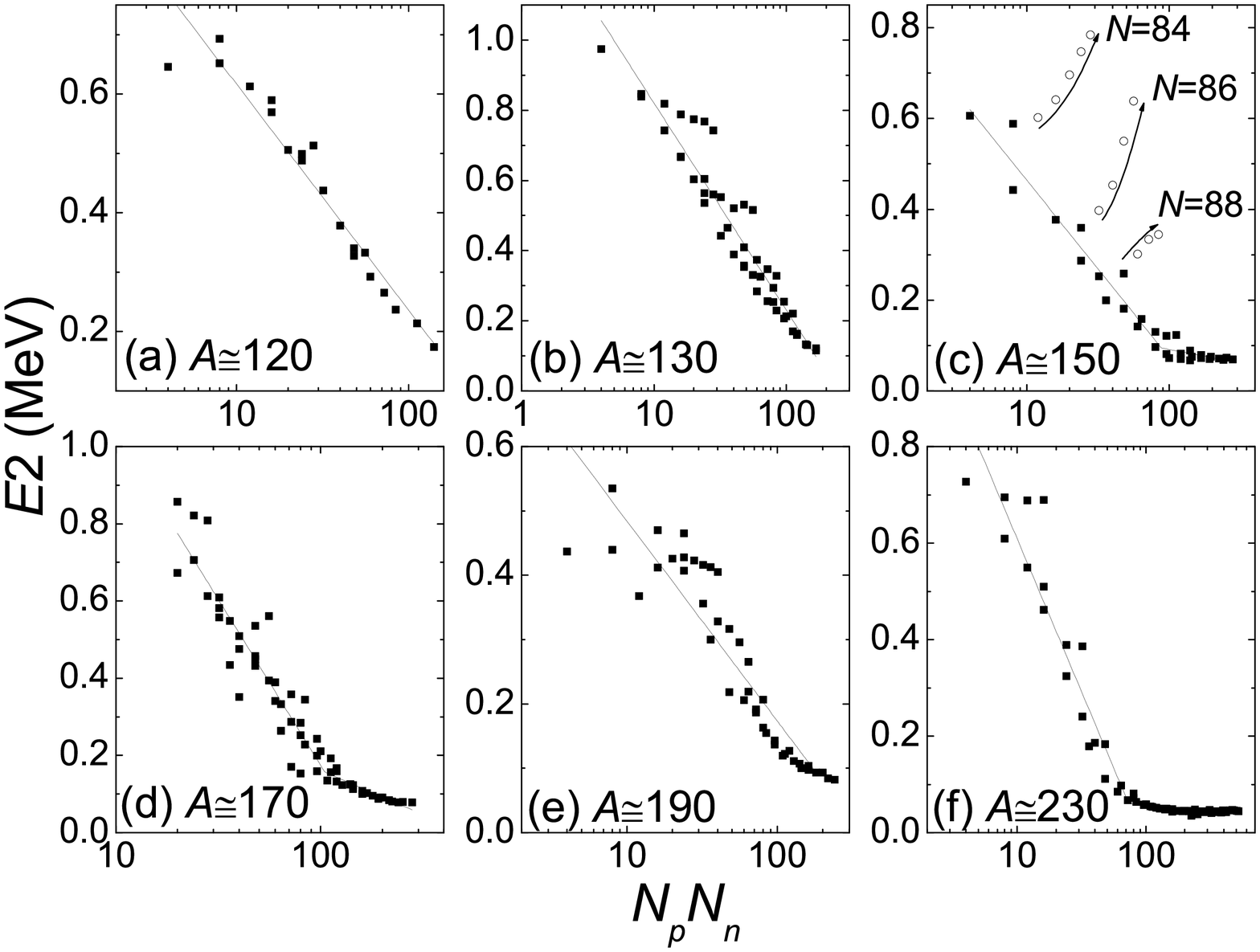}
\caption{$E2$ values plotted versus the logarithmically-scaled $N_pN_n$ product. All the data are from the ENSDF \cite{ensdf} compilation. Solid lines represent the results of fitting to Eqs. (\ref{eq-linear}) and (\ref{eq-bilinear}) with the best-fit results listed in Table \ref{tb-e2}. Three abnormal branches due to the $Z=64$ subshell are highlighted by circles in (c).}\label{fig-e2}
\end{figure}

As expected, Fig. \ref{fig-e2} exhibits a linear behavior of $E2$ values against $\ln(N_pN_n)$ in all of the regions, except for the $N=84$, 86 and 88 isotones around $A\simeq150$. The $N=84-88$ anomaly in the $N_pN_n$ scheme has been attributed to the role of the $Z=64$ subshell \cite{z64-1,z64-2}. If we exclude this anomalous data in Fig. \ref{fig-e2}(c), linearity also emerges for the $A\simeq 150$ region. This indicates that the constraint of Eq. (\ref{eq-linear}) is indeed of general relevance when treating $E2$ values in the $N_pN_n$ scheme. We also note that the slope of the $E2$ linearity  dramatically changes around some critical point in the $A>140$ regions, corresponding to the known $E2$ saturation \cite{zhao-npnn-fit}. To quantitatively determine the critical point, we have performed a bilinear fit for the  $E2$ vs $N_pN_n$ plots in the $A>140$ regions. The corresponding fitting function is defined as
\begin{equation}\label{eq-bilinear}
F=
\begin{cases}
c_1 \ln (N_pN_n)+c_2 & \text{for $N_pN_n<(N_pN_n)_c$}\\
c_3 \ln (N_pN_n)+c_4 & \text{for $N_pN_n>(N_pN_n)_c$}
\end{cases}
,
\end{equation}
where $c_1$, $c_2$, $c_3$ and $(N_pN_n)_c$ are fitting variables, and $c_4=c_1\ln (N_pN_n)_c+c_2-c_3\ln (N_pN_n)_c$ to keep functional continuity. Note that the subscript $c$ refers to the critical value of $N_pN_n$. $E2$ values in $A\simeq 120$ and 130 do not reach saturation. Thus, in Figs. \ref{fig-e2}(a) and (b) we perform single-segment linear fits to Eq. (\ref{eq-linear}) with $c_1$ and $c_2$ as fitting variables. 

The best-fit results are listed in Table \ref{tb-e2}, and the corresponding linear fits are illustrated in Fig. \ref{fig-e2} by solid lines. One sees that the linear fits reasonably describe the tendency of $E2$ values, further confirming the general validity of Eq. (\ref{eq-linear}). In Table \ref{tb-e2}, all the $A>140$ regions have $c_3\simeq 0$ and $(N_pN_n)_c\simeq 90$ within fitting errors. $c_3\simeq 0$ is a natural result of $E2$ saturation noted above. The rough uniformity of the $(N_pN_n)_c\simeq 90$ critical point may be explained by the Federman-Pittel mechanism \cite{federman-pittel}, which emphasized that nuclear deformation, which may be empirically represented by $E2$ values, is mostly governed by the $pn$ interaction between orbits with the same orbital angular momentum (spin-orbit partners), {\it e.g.,} $1g_{9/2}-1g_{7/2}$, $1h_{11/2}-1h_{9/2}$, and $1i_{13/2}-1i_{11/2}$. The occupation-number limits for all of these orbits are near 10. Thus, the $(N_pN_n)_c\simeq 90$ critical point seems to correspond to almost full occupation of the relevant spin-orbit partners. For nuclei with $N_pN_n> 90$, the Pauli principle prevents additional valence nucleons/holes from occupying the spin-orbit partners, and thus these nucleons contribute little to the deformation. As a result, the $E2$ value saturates.

\begin{table}
\caption{Best-fit results from the $E2$ vs $N_pN_n$ plots in Fig. \ref{fig-e2}; see Eqs. (\ref{eq-linear}) and (\ref{eq-bilinear}) for definitions.}\label{tb-e2}
\begin{tabular}{cccccc}
\hline\hline
       &    $c_1$     &    $c_2$     &    $c_3$     &    $(N_pN_n)_c$     \\
\hline
$A\simeq 120$ &    -0.38(2)      &    1.00(3) &           &           \\
$A\simeq 130$ &    -0.59(2)      &    1.41(4) &           &           \\
$A\simeq 150$ &    -0.170(9)    &    0.85(3) &    -0.03(2)      &    $88\pm 12$      \\
$A\simeq 170$ &    -0.37(2)      &    1.89(8) &    -0.10(4)      &    $105\pm 11$    \\
$A\simeq 190$ &    -0.14(1)      &    0.80(4) &    -0.0(3)  &    $178\pm105$  \\
$A\simeq 230$ &    -0.28(1)      &    1.24(4) &    -0.01(1)      &    $73\pm 7$ \\
\hline\hline
\end{tabular}
\end{table}

\subsection{$B(E2)$}
\begin{figure}
\includegraphics[angle=0,width=0.48\textwidth]{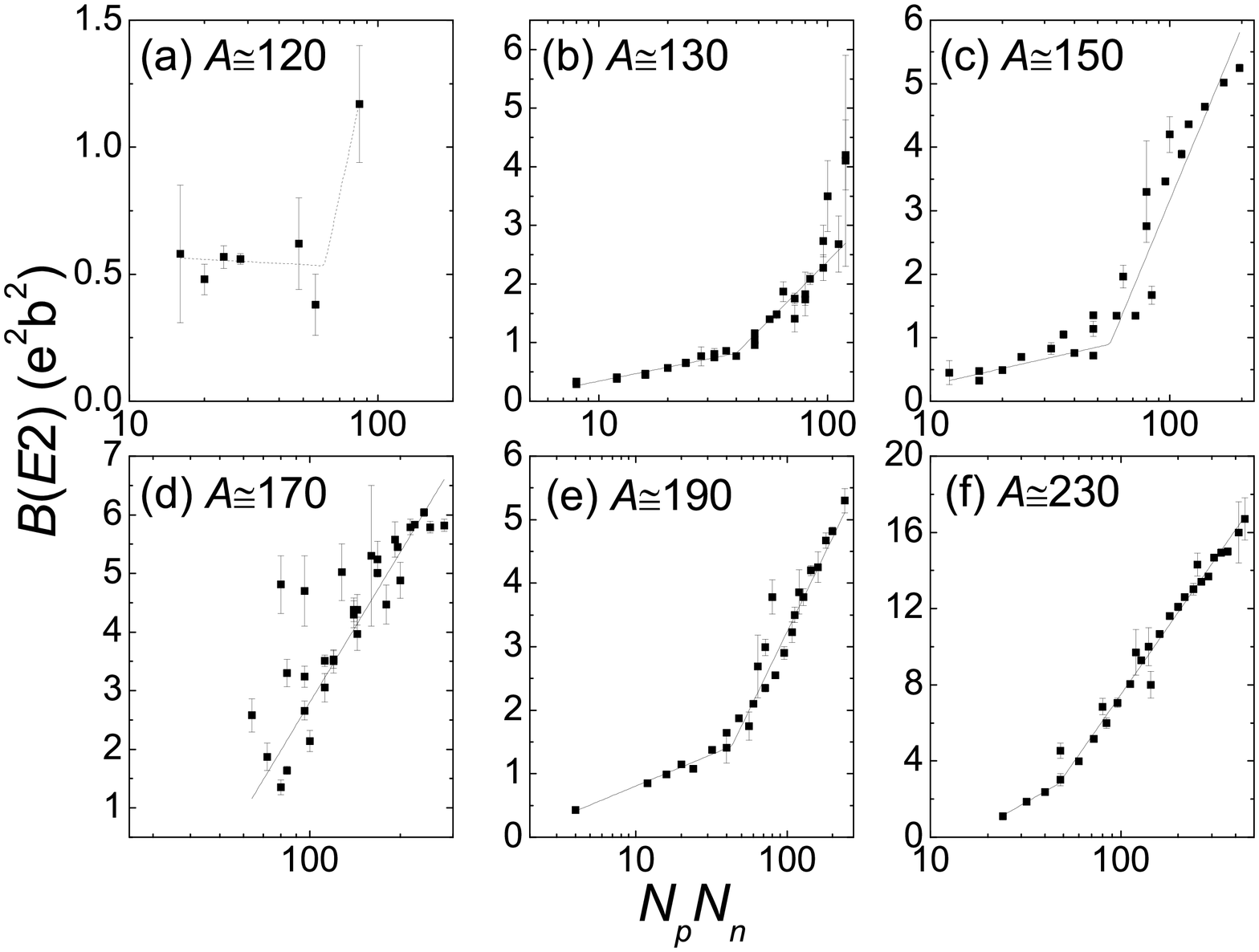}
\caption{$B(E2)$ against the logarithmically scaled $N_pN_n$ product. All the data are from Ref. \cite{be2-raman}. Solid lines demonstrate the linear relation with best-fit variables listed in Table \ref{tb-be2}. The fitting of (a) is not convergent due to lack of data. Still, we present this divergent fitting result with a dot line as a guide to the eye.}\label{fig-be2}
\end{figure}

In Fig. \ref{fig-be2}, the $B(E2)$  values are plotted against the logarithmacally-scaled $N_pN_n$ values for the same mass regions as were used for E2 values. Here too a roughly linear behavior emerges in most of the regions considered, as expected.  Another $N_pN_n$ critical point of the $B(E2)$ evolutions is also evident, across which the evolution of $B(E2)$ values exhibits slopes that are clearly enlarged.  Again, we have performed a bilinear fit for the  $E2$ vs $N_pN_n$ plots with Eq. (\ref{eq-bilinear}), but now omitting the $A\simeq 120$ and $170$ regions. Due to a lack of experimental data, the fit for the $A\simeq 120$ region is not convergent. In the $A\simeq 170$ region, experimental $B(E2)$ values with $N_pN_n$ smaller than the critical value cannot be determined, since the corresponding nuclei are all near $^{164}$Pb and thus beyond the proton drop line. Thus, for the $B(E2)$ systematics in the $A\simeq 170$ region we only use a single-segment linear fit of Eq. (\ref{eq-linear}).

\begin{table}
\caption{Best-fit results of $B(E2)$ vs $N_pN_n$ plots shown in Fig. \ref{fig-e2} (see Eqs. (\ref{eq-linear}) and (\ref{eq-bilinear}) for definitions). The fit for the $A\simeq 120$ region is not convergent, and is thus omitted here.}\label{tb-be2}
\begin{tabular}{cccccc}
\hline\hline
       &    $c_1$     &    $c_2$     &    $c_3$     &    $(N_pN_n)_c$     \\
\hline
$A\simeq 130$ &    0.34(1) &    -0.45(4)      &    1.7(2)   &    $40\pm 2$ \\
$A\simeq 150$ &    0.4(1)   &    -0.6(4)  &    3.9(3)   &    $56\pm 4$ \\
$A\simeq 170$ &    8.5(4)   &    -14(1)  &           &           \\
$A\simeq 190$ &    0.43(2) &    -0.18(5)      &    2.1(2)   &    $43\pm 3$ \\
$A\simeq 230$ &    2.5(9)   &    -7(3)     &    6.2(1)   &    $47\pm 5$ \\
\hline\hline
\end{tabular}
\end{table}

We illustrate the best linear fits with solid lines in Fig. \ref{fig-be2} and list the corresponding best-fit parameters in Table \ref{tb-be2}. As is clear from the table, the best-fit $N_pN_n$ critical points associated with $B(E2)$ values all cluster around $(N_pN_n)_c\simeq 45$, very different from those that emerged in the treatment of $E2$ values (see Table \ref{tb-e2}).  The $E2$ critical point corresponds to its saturation. In contrast,  the $B(E2)$ values do not saturate near their critical value, but rather continue to increase, albeit with a more pronounced slope.  It would seem therefore that the $(N_pN_n)_c$ of $B(E2)$ corresponds to an underlying transition in the nature of collective motion, rather than to saturation. The detailed mechanism that gives rise to this critical point still requires further investigation. The fact that the critical value of $N_pN_n$ associated with $B(E2)$ values is half of that for $E2$ values might provide a useful clue to its origin.

\section{Decoupling of $Q_{\alpha}$, $S_n$ and $S_p$}
\begin{figure*}
\includegraphics[angle=0,width=0.9\textwidth]{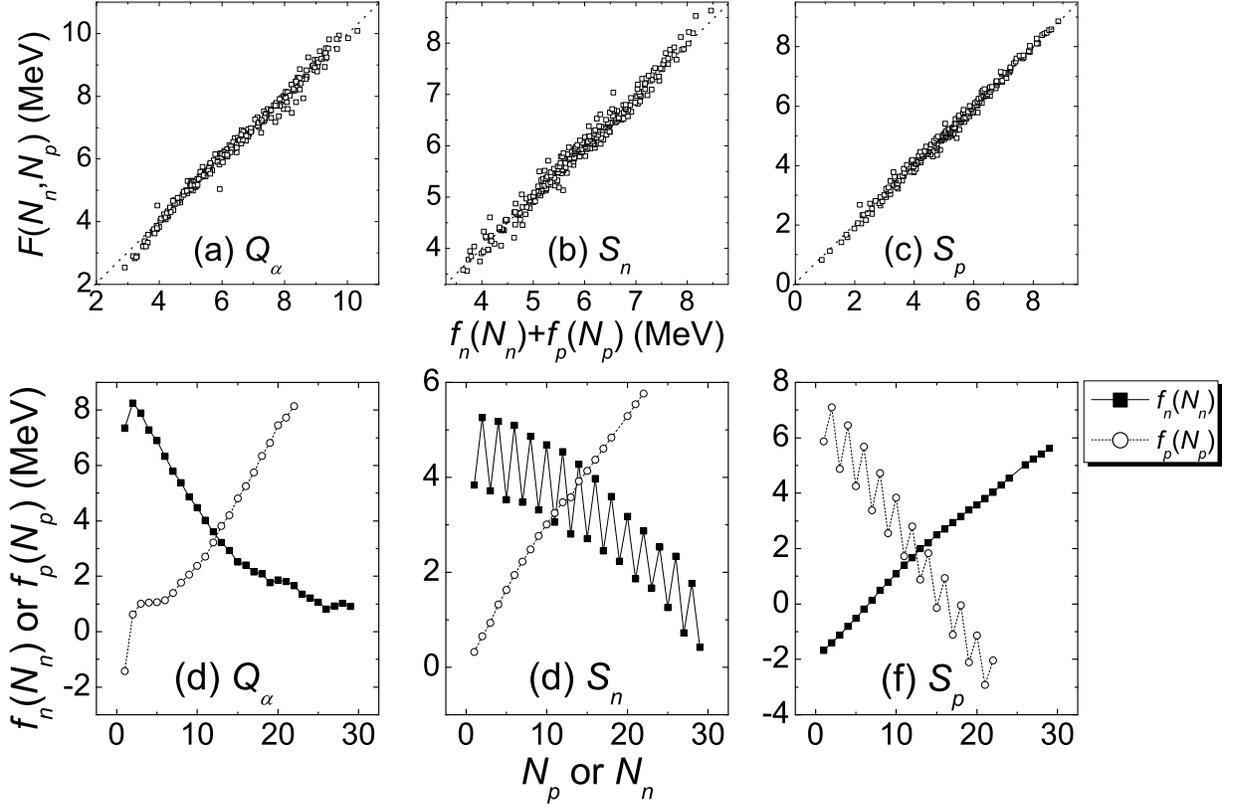}
\caption{Final results of the decoupling analysis of the $Q_{\alpha}$, $S_n$ and $S_p$ experimental data in $82<Z\leq104,~126<N\leq 155$ region. The data are extracted from the AME2012 \cite{q-mass} mass table. In (a), (b) and (c), the experimental $Q_{\alpha}$, $S_n$ and $S_p$ values are plotted against the decoupled $f_p(N_p)+f_N(N_n)$, respectively. The diagonal dotted lines in (a), (b) and (c) correspond to the exact $F(N_p,N_n)=f_p(N_p)+f_n(N_n)$ relation. Panels (d), (e) and (f) illustrate the evolution of the decoupled $f_p(N_p)$ and $f_n(N_n)$ results for $Q_{\alpha}$, $S_n$ and $S_p$, respectively. The results exhibit smooth regional systematics with the expected odd-even staggering for $S_n$ and $S_p$.}\label{fig-decouple}
\end{figure*}

Since the quantities $Q_{\alpha}$, $S_n$ and $S_p$ are all governed by Eq. (\ref{local}), they can be decoupled as in Eq. (\ref{formal}). Because they all vary smoothly, despite the odd-even staggering of $S_n$ and $S_p$, some regional systematics may be examined via these decoupled results. To accomplish this, we carry out a $\chi^2$ fit of Eq. (\ref{formal}) to decouple the experimental $Q_{\alpha}$, $S_n$ and $S_p$ data. The analysis is carried out for nuclei in the $82<Z\leq104,~126<N\leq 155$ region. All of the experimental data are extracted from the AME2012 mass table \cite{q-mass}. Details on the decoupling procedure are described in the Appendix. The final results of the decoupling analysis are presented in Fig. \ref{fig-decouple}.

According to Figs. \ref{fig-decouple}(a), (b) and (c), our decoupled $f_p(N_p)+f_n(N_n)$ values fit well to the $F(N_p,N_n)$ values from experiment for all of the nuclear observables under investigation, thereby demonstrating the validity of the decoupling scheme based on Eq. (\ref{formal}). The regional systematics for $f_p(N_p)$ and $f_n(N_n)$ in Figs. \ref{fig-decouple} (d), (e) and (f) are evident.

In Fig. \ref{fig-decouple}(d), the $Q_{\alpha}$ values decrease with increasing $N_n$ and decreasing $N_p$, {\it i.e.}, ${\rm d} f_n/{\rm d} N_n<0$ and ${\rm d} f_p/{\rm d} N_p>0$. This can be attributed to the negative effect of the Coulomb and symmetry energies on nuclear binding as follows. The nuclear binding energy of the Bethe-Weizsacker formula \cite{bw-1,bw-2} is given by
\begin{equation}\label{eq-bw}
\begin{aligned}
B(N,Z)=&a_vA-a_sA^{2/3}-a_cZ^2A^{-1/3}\\
&-a_I\left(\frac{A}{2}-Z\right)^2A^{-1}+a_p\delta A^{-1/2},
\end{aligned}
\end{equation}
where $a_v$, $a_s$, $a_c$, $a_I$, $a_p$ are parameters associated with the volume term, the surface term, the Coulomb energy, the symmetry energy, and the pairing energy, respectively. From this, $Q_{\alpha}$ can be expressed as
\begin{equation}\label{eq-qa}
\begin{aligned}
Q_{\alpha}=&B_{\alpha}-\left[B(N,Z)-B(N-2,Z-2)\right]\\
=&B_{\alpha}-4a_v+\frac{8a_s}{3A^{1/3}}+4a_c\frac{Z}{A^{1/3}}\left(1-\frac{Z}{3A}\right)\\
&-a_I\left(\frac{N-Z}{A}\right)^2,
\end{aligned}
\end{equation}
where the pairing energies approximately cancel each other for heavy nuclei, and $B_{\alpha}$ is the binding energy of the $\alpha$ particle. The $1/A$ and $1/A^{1/3}$ term should vary slowly for heavy nuclei. Thus, we assume them to be constant, so that derivatives of $Q_{\alpha}$ become simplified as
\begin{equation}\label{eq-pq}
\begin{aligned}
\left(\frac{\partial Q_{\alpha}}{\partial N}\right)_Z=&\frac{{\rm d} f_n}{{\rm d} N_n}\simeq-2a_I\frac{N-Z}{A^2}\\
\left(\frac{\partial Q_{\alpha}}{\partial Z}\right)_N=&\frac{{\rm d} f_p}{{\rm d} N_p}\simeq\frac{4a_c}{A^{1/3}}\left(1-\frac{2Z}{3A}\right)+2a_I\frac{N-Z}{A^2}
\end{aligned}.
\end{equation}
The Coulomb and symmetry energies decrease nuclear stability, {\it i.e.}, $a_c$ and $a_I$ are positive, which leads to ${\rm d} f_n/{\rm d} N_n<0$ and ${\rm d} f_p/{\rm d} N_p>0$, given that $N>Z$ for heavy nuclei. This explains the observed tendencies exhibited by $f_n(N_n)$ and $f_p(N_p)$ in Fig. \ref{fig-decouple}(d). We note that because of the Coulomb energy, {\it i.e.}, the first term of $(\partial Q_{\alpha}/\partial Z)_N$ in Eq. (\ref{eq-pq}), ${\rm d} f_p/{\rm d} N_p$ always has a larger magnitude than ${\rm d} f_n/{\rm d} N_n$. As a result, the $f_p(N_p)$ evolution of $Q_{\alpha}$ is sharper than that of $f_n(N_n)$, as observed in Fig. \ref{fig-decouple}(d).

In Figs. \ref{fig-decouple}(e) and (f), the odd-even staggering of $f_n(N_n)$/$f_p(N_p)$ for $S_n$/$S_p$ is observed clearly, and corresponds to the effect of pairing between like nucleons. By smoothing the odd-even staggering, $S_n$/$S_p$ generally decreases with increasing $N_n$/$N_p$, and increases with increasing $N_p$/$N_n$, implying that the non-pairing interaction between like nucleons is repulsive and that the $pn$ interaction is attractive.

We also note that the observed evolution of $S_n$ and $S_p$ in Figs. \ref{fig-decouple}(e) and (f) agrees with their previously proposed linear systematics with respect to the $Z/N$ and $N/Z$ ratios \cite{snsp-vogt}. We adopt empirical formulas from Ref. \cite{snsp-vogt} to compare the sharpness of the evolution of $S_n$ and $S_p$ as follows:
\begin{equation}
\begin{aligned}
S_n=&a\frac{Z}{N}+b~,\\
S_p=&a\frac{N}{Z}+b-a_cZA^{-1/3}~,
\end{aligned}
\end{equation}
where $a$ and $b$ are constants within a major shell, the $a_cZA^{-1/3}$ term comes from the Coulomb energy, and the pairing term is neglected here to smooth the odd-even staggering. Thus,
\begin{equation}
\begin{aligned}
&\left.
\begin{aligned}
\left(\frac{\partial S_n}{\partial N}\right)_Z=&\frac{{\rm d} f_n}{{\rm d} N_n}=-a\frac{Z}{N^2}\\
\left(\frac{\partial S_n}{\partial Z}\right)_N=&\frac{{\rm d} f_p}{{\rm d} N_p}=a\frac{1}{N}
\end{aligned}
~~~~~~~~~~~~~\right\}&{\rm for}~S_n\\
&\left.
\begin{aligned}
\left(\frac{\partial S_p}{\partial N}\right)_Z=&\frac{{\rm d} f_n}{{\rm d} N_n}=a\frac{1}{Z}\\
\left(\frac{\partial S_p}{\partial Z}\right)_N=&\frac{{\rm d} f_p}{{\rm d} N_p}=-a\frac{N}{Z^2}-a_cA^{-1/3}
\end{aligned}
\right\}&{\rm for}~S_p
\end{aligned}.
\end{equation}
According to the analysis in Ref. \cite{snsp-vogt}, $a$ and $a_c$ are always positive. Thus,
\begin{equation}
\left|\frac{{\rm d} f_n}{{\rm d} N_n}\right|<\left|\frac{{\rm d} f_p}{{\rm d} N_p}\right|
\end{equation}
for both $S_n$ and $S_p$, indicating that the nucleon separation energy is always more sensitive to the proton number, as illustrated in Figs. \ref{fig-decouple}(e) and (f).

\section{summary}\label{sum}
To summarize, we have studied the regional systematics of $E2$, $B(E2)$, $Q_{\alpha}$, $S_n$ and $S_p$ values based on their local correlations, as defined by Eq. (\ref{local}). Constrained by such local correlations, $N_pN_n$ plots of $E2$ and $B(E2)$ should and indeed do present robust linearity in the logarithmic scale. Such a linear behavior is adopted to quantitatively probe the saturation of $E2$ in the vicinity of $N_pN_n\sim 90$, which was then explained using the Federman-Pittel mechanism. A new and unified critical point of $B(E2)$ evolution is identified around $N_pN_n\sim 45$, which we believe deserves further clarification. Using the decoupling scheme of Eq. (\ref{formal}), as derived from the generalization of Eq. (\ref{local}), we then extracted the proton and neutron contributions to the experimental $Q_{\alpha}$, $S_n$ and $S_p$ values. These decoupled results exhibit smooth regional systematics beyond the $N_pN_n$ scheme. Such regional systematics agree with previous empirical models, suggesting that {\it the decoupling scheme is a practical way to study regional evolution of non-$N_pN_n$ systematized nuclear observables that follow Eq (\ref{local}).} In closing, the results presented here suggest that local correlations may provide a new and perhaps clearer vision of nuclear regional evolution.

\acknowledgements
We thank Prof. Y. M. Zhao for fruitful discussions, and Prof. S. Pittel for his careful proofreading. This work was supported by the National Natural Science Foundation of China under Grant Nos. 11647059, 11305151, 11225524, 11675101, the Research Fund for the Doctoral Program of the Southwest University of Science and Technology under Grant No. 14zx7102, and the Graduate Education Reform Project of the Southwest University of Science and Technology under Grant No. 17sxb119.

\appendix*
\section{Decoupling process}\label{dec}
We adopt a $\chi^2$ fitting of $F(N_p,N_n)=f_p(N_p)+f_n(N_n)$, with $f_p(N_p)$ and $f_n(N_n)$ as fitting parameters, to decouple the $F(N_p,N_n)$ values that come from experiment. To simplify our description, we denote the number of $N_p$ and $N_n$ values under investigation as $\Lambda_{\pi}$ and $\Lambda_{\nu}$, respectively. Thus, the variables to be fitted are $\Lambda_{\pi}$ $f_p(N_p)$ and $\Lambda_{\nu}$ $f_n(N_n)$.

We note that if a pair of $f_p(N_p)$ and $f_n(N_n)$ variables satisfies the $F(N_p,N_n)=f_p(N_p)+f_n(N_n)$ relation, another pair $f_p(N_p)+C$ and $f_n(N_n)-C$ also does, with an arbitrary constant $C$. To remove this arbitrariness, and to ensure that $f_p(N_p)$ and $f_n(N_n)$ have the same order of magnitude, we further require
\begin{equation}\label{con-fit}
\frac{\sum\limits_{N_p}f_p(N_p)}{\Lambda_{\pi}}=\frac{\sum\limits_{N_n}f_n(N_n)}{\Lambda_{\nu}}.
\end{equation}

We define our $\chi^2$ function as
\begin{equation}
\chi^2=\sum\limits_{N_p,N_n}\Bigg\{F(N_p,N_n)-f_p(N_p)-f_n(N_n)\Bigg\}^2.
\end{equation}
The $\chi^2$ minimum under the constraint of Eq. (\ref{con-fit}) provides the best fit of $F(N_p,N_n)=f_p(N_p)+f_n(N_n)$. To reach this minimum, we introduce the Lagrangian
\begin{equation}
\mathcal{L}=\chi^2+\lambda\left\{\Lambda_{\nu}\sum\limits_{N_p}f_p(N_p)-\Lambda_{\pi}\sum\limits_{N_n}f_n(N_n)\right\},
\end{equation}
with $\lambda$ as a Lagrange multiplier. The solution of the set of partial differential equations,
\begin{equation}\label{linear-fit}
\frac{\partial \mathcal{L}}{\partial f_p(N_p)}=0,~
\frac{\partial \mathcal{L}}{\partial f_n(N_n)}=0,~
\frac{\partial \mathcal{L}}{\partial \lambda}=0,
\end{equation}
corresponds to the desired $\chi^2$ minimum, {\it i.e.}, our decoupling result.

\end{document}